\documentclass[twocolumn,showpacs,preprintnumbers,amsmath,amssymb]{revtex4}

\usepackage{graphicx}
\usepackage{dcolumn}
\usepackage{bm}
\usepackage{upgreek}
\long\def\symbolfootnote#1{\begingroup\def\thefootnote{a)}\footnote{#1}\endgroup}

\begin{document}
%
%
\title{Opto-fluidic third order distributed feed-back dye laser}
\author{Morten Gersborg-Hansen}
\author{Anders Kristensen\symbolfootnote{Electronic address: ak@mic.dtu.dk; URL: www.mic.dtu.dk/nil}}

\affiliation{%
MIC -- Department of Micro and Nanotechnology, NanoDTU, Technical
University of Denmark, Building 345 east, \O rsteds Plads, DK-2800
Kongens Lyngby, Denmark
}%
\date{\today}
\begin{abstract}
This letter describes the design and operation of a polymer-based
third order distributed feed-back (DFB) microfluidic dye laser.
The device relies on light-confinement in a nano-structured
polymer film where an array of nanofluidic channels is filled by
capillary action with a liquid dye solution which has a refractive
index lower than that of the polymer. In combination with a third
order DFB grating, formed by the array of nanofluidic channels,
this yields a low threshold for lasing. The laser is
straight-forward to integrate on Lab-on-a-Chip micro-systems where
coherent, tunable light in the visible range is desired.
\\\\
\end{abstract}
\maketitle

Integration of optical and fluidic functionalities on a chip has
recently been investigated by several research groups to realize
novel opto-fluidic laser devices, microfluidic dye lasers,
suitable for Lab-on-a-Chip
micro-systems~\cite{helbo2003,balslev05optexp,vezenov05,li06optexp}.
These optically pumped devices consist of microfluidic channels
with an embedded optical resonator and a liquid laser dye is used
as active gain medium. In general, these devices can be added to a
microfluidic chip without adding additional fabrication
steps~\cite{balslev06labchip}, thus offering a simple way of
integrating optical transducers to Lab-on-a-Chip
micro-systems~\cite{verpoorte2003}, and pave the way for advanced
integrated sensor concepts~\cite{lading03spie}.

The threshold for lasing is a key parameter for the feasibility of
these devices in a future technology. Distributed feed-back (DFB)
laser resonators have proven particularly suited in this respect.
A DFB microfluidic dye laser was first demonstrated by Balslev and
Kristensen~\cite{balslev05optexp}, who used a high order Bragg
grating in an $8~\upmu\rm{m}$ thick polymer film to obtain
feed-back. Single mode lasing with a threshold fluence of
approximately $20~\upmu\rm{J}/\rm{mm}^2$ was obtained due to mode
selective losses in the multi-mode structure where light was not
guided in the fluidic segments. Vezenov \emph{et
al.}~\cite{vezenov05} obtained liquid-core waveguiding using a low
refractive index polymer and a high refractive index liquid. Li
\emph{et al.}~\cite{li06optexp} exploited this concept to realize
a 15th order DFB laser with a record-low threshold fluence for
lasing of approximately $8~\upmu\rm{J}/\rm{mm}^2$.

However, the 15th order Bragg grating still gives rise to
scattering of light out of the chip plane~\cite{hunsperger02}.
These losses can be reduced by employing a lower order Bragg
grating. Ideally, a first order Bragg grating combined with planar
waveguiding in the polymer film could be used. The requirement of
a high refractive index liquid and low refractive index
polymer~\cite{vezenov05,li06optexp} may be relaxed by reducing the
dimensions of the resonator segments as the sub-wavelength regime
is entered~\cite{lalanne03encyclopteng}.

\begin{figure}[b]
  \centering
  \includegraphics[width=8cm]{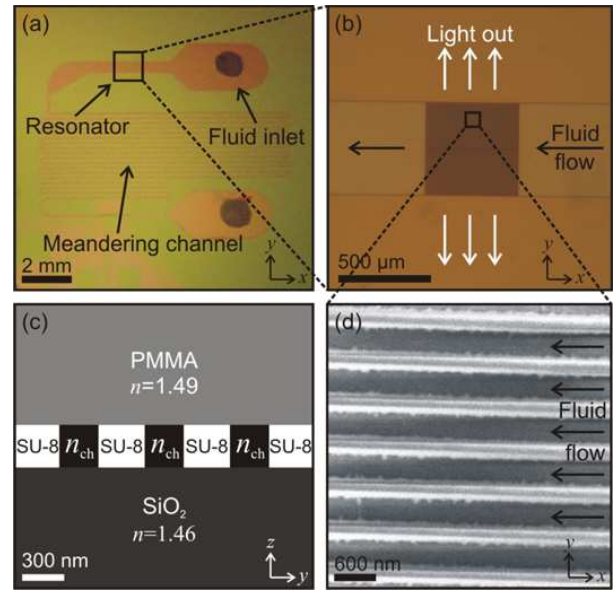}
  \caption{Overview of the fabricated microfluidic dye laser.
  (a) Top-view picture of the over-all layout of the
  fabricated chip. The shallow 300 nm high meandering
channel facilitates capillary filling of the
  embedded laser resonator. (b) Optical micrograph of the DFB
   laser resonator embedded in the shallow meandering channel.
   (c) Side-view sketch showing the layers of the DFB laser resonator.
   (d) Scanning
  electron micrograph showing the third order Bragg grating of nanofluidic channels
 which constitutes the DFB laser resonator (prior to bonding).
  }\label{fig:chip}
\end{figure}

In this Letter, we present the design and operation of a
polymer-based third order DFB microfluidic dye laser. The device
is fabricated by a flexible lithography technique, combining
electron beam lithography (EBL) and UV lithography (UVL) in a
single polymer film, enabling fast prototyping of device designs.
In the device, see Fig.~1, a third order DFB resonator is formed
by infiltrating a periodic array of nanofluidic channels with a
liquid laser dye. The filling is mediated by capillary action,
thus simplifying the operation of the device as opposed to
conventional setups with external syringe pumps. Replenishing of
the dye in the cavity region is achieved for hours through
capillary filling of a $16~\rm{cm}$ long 100 $\upmu$m wide
meandering channel following the cavity region. The speed of the
advancing fluid front in the meandering channel decreases with
time~\cite{tas04}. The dye solution in the cavity region is
replaced in 30 s -- 5 min. during the first 30 min. of experiments
(ethylene glycol solution). The flow rate in the 500 $\upmu$m
$\times$ 500 $\upmu$m cavity region can be further increased by
increasing the width of the meandering channel. By employing a
third order Bragg grating, we demonstrate lasing with a threshold
fluence comparable to the results of \cite{li06optexp}. The
sub-wavelength dimensions of the third order DFB grating yield a
low coupling loss for the light when traversing the dye filled
nanofluidic channels, and combined with the feed-back of the DFB
grating, this yields an efficient laser device.

The laser is based on a planar waveguide structure supporting a
single propagating TE-TM mode. The basic waveguide structure
consists of a SiO$_2$ (refractive index $n=1.46$) buffer
substrate, a 300 nm thick core layer of the negative-tone resist
SU-8 ($n=1.59$) and a top cladding of poly-methylmethacrylate
(PMMA) ($n=1.49$). The device structure is formed by nanofluidic
channels defined lithographically in the SU-8 film. The laser
resonator consists of an array of 300 nm high nanochannels of
period $\Lambda=601~\rm{nm}$ which comprises a third order Bragg
grating with a central $\pi/2$ phase-shift, embedded in a
$500~\upmu\rm{m}$ wide shallow nanochannel.

To fabricate the devices, a 430 nm thick film of SU-8 2000
resist~\cite{microchem} is spin-coated onto a Si substrate with a
$2.5~\upmu\rm{m}$ thick thermally grown oxide layer. The wafer is
baked at 90$^\circ$C for 1 min. The device structure is defined by
combined EBL and UVL in the SU-8 film. The nano-structures of the
third order DFB grating are defined by 100 kV electron beam
exposure (JEOL-JBX9300FS, dose $3~\upmu\rm{C}/\rm{cm}^2$). The
total writing time for all 32 devices on a 4" wafer is 15 min.
After electron beam exposure, the micron-sized structures
(meandering channel and reservoirs) are defined in the same
polymer film by UV exposure (20 s at $8.9~\rm{mW}/\rm{cm}^2$), the
wafer is post-exposure baked at 90$^\circ$C for 1 min., and the
micro and nano-structures are developed simultaneously in
propylene glycol monomethyl ether acetate (PGMEA) for 30 s,
followed by an iso-propyl alcohol (IPA) rinse. The wafer is
subsequently subjected to a 20 s soft oxygen
plasma~\cite{bilenberg2006mee} to remove any residues of SU-8 at
the SiO$_2$ surface in the nano-channels of the DFB resonator.
After the plasma treatment, the structures are 300 nm high, and
the heights of the EBL and UVL defined structures are matched
within 30 nm. Fluidic access holes are formed by micro powder
blasting~\cite{danville} and the channels are sealed by a glass
lid, using adhesive bonding by means of a $5~\upmu\rm{m}$ thick
PMMA film (2200 N, 140$^\circ$C, 10
min.)~\cite{bilenberg2004bond}.

The laser is characterized by adding a droplet of dye solution to
the fluid inlet, thus filling the DFB resonator structure by
capillary action. In this Letter, the laser dye Rhodamine 6G (R6G)
is dissolved in ethylene glycol ($n=1.43$), ethanol ($n=1.33$),
and in a 1:1 mixture of ethylene glycol and benzyl alcohol
($n=1.485$ by experimentally confirmed linear extrapolation of the
refractive indices), all with a concentration of
$2\times10^{-2}~\rm{mol/L}$ and a refractive index below that of
the polymer surrounding the channel. The laser is optically pumped
through the glass lid by a frequency doubled Nd:YAG laser at 532
nm (5 ns pulse duration, 10 Hz repetition rate).

\begin{figure}[b]
  \centering
  \includegraphics[width=8cm]{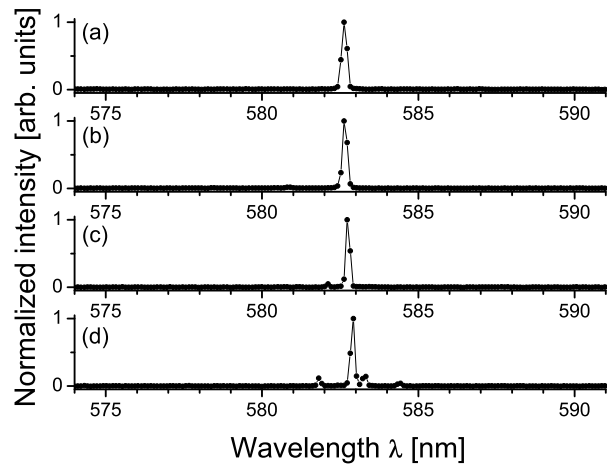}
  \caption{Normalized laser spectra from four nominally identical
chips
  demonstrating wafer-scale spectral reproducibility of the laser. On each chip, the
  DFB laser resonator is filled by capillary action
  with Rhodamine 6G (R6G)
  dissolved in ethylene glycol. The average laser wavelength
  of the chips is 582.72 nm with a standard deviation of 0.14
  nm.}\label{fig:eg-chips}
\end{figure}

Fig.~2 shows laser spectra from four different chips filled with
R6G in ethylene glycol. The results demonstrate wafer-scale
spectral reproducibility of the laser, exhibiting narrow-linewidth
emission, polarized perpendicularly to the chip plane (TM). The
average laser wavelength is 582.72 nm with a standard deviation of
0.14 nm, at the resolution limit of our spectrometer (0.15 nm).
The deviation and the minor peaks in Fig.~2(c),(d) may arise from
grating imperfections due to fabrication defects. Due to limited
spectrometer resolution, our equipment does not allow us to
determine whether the laser is truly operating in a single mode,
and the spectral shape and linewidth of the laser remain
undetermined.

\begin{figure}[htb]
  \centering
  \includegraphics[width=8cm]{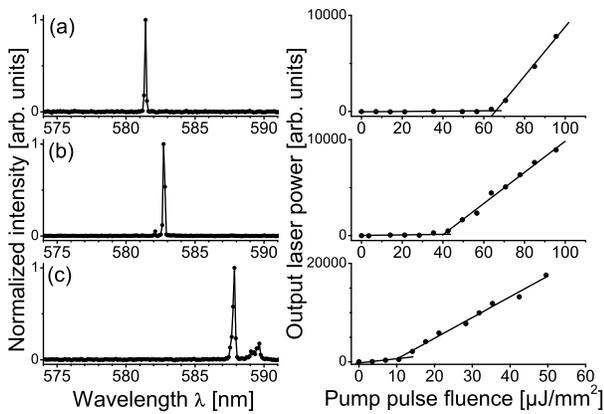}
  \caption{Normalized laser spectra and corresponding laser output power curves as a
  function of pump pulse fluence for R6G in three different
  solutions:
  (a) R6G in ethanol.
  (b) R6G in ethylene glycol.
  (c) R6G in a 1:1 mixture of ethylene glycol and benzyl alcohol.
  The results are summarized in Table~\ref{tab:summary}.
  }\label{fig:threshold}
\end{figure}

Fig.~3 shows laser spectra from chips with R6G dissolved in
different solvents, thus demonstrating tunability of the laser
from 581.41 nm to 587.86 nm. A quantitative theoretical prediction
of the shift in wavelength requires a full solution of Maxwell's
equations in the fabricated structure, since the structures are of
sub-wavelength dimensions~\cite{lalanne03encyclopteng}. In
Fig.~3(c), the minor side-mode shifted 1.8~nm from the main peak
may be attributed to an additional phase-shift introduced in the
DFB grating by a stitching error during electron beam exposure.
The dye laser output power is found by integrating the measured
laser spectrum for each value of the average pump pulse fluence.
The dye laser output power vs. average pump pulse fluence follows
the standard pump/output relation of two linear segments around a
lasing threshold fluence $Q_{\rm{th}}$.

The coupling loss $\gamma$ for a period of the DFB grating has
been modelled by a finite-difference beam propagation
method~\cite{balslev06optexp,hoekstra92}. Light of the observed
wavelength travelling in the TM mode of the single mode polymer
waveguide is propagated through a 273 nm wide channel filled with
a liquid of refractive index $n_{\rm{ch}}$, corresponding to the
value of the used solvent. The calculations yield the energy
propagation loss corresponding to the coupling loss for a period
of the DFB grating. The loss decreases for increasing
$n_{\rm{ch}}$, see Table~\ref{tab:summary}. To obtain lasing in
the resonator, the amplification of the R6G must equal the loss.
As expected, the experiments show that the threshold fluence for
lasing decreases as the refractive index of the R6G solution
increases, see Table~\ref{tab:summary}. Further, when considering
the absorbed pump energy throughout the device, the effective
threshold fluence may be considerably lower.

In summary, we have demonstrated a third order DFB microfluidic
dye laser with a low threshold for lasing and operating with
liquids of refractive index below the refractive indices of the
surrounding polymer. Although experimental uncertainties make a
direct comparison of lasing thresholds difficult, our results
imply the importance of both liquid core waveguiding
\cite{li06optexp} and low order Bragg reflection. The device
relies on light-confinement in a nano-structured polymer film
where the individual resonator elements -- nanofluidic channels
and polymer walls -- are of sub-wavelength dimensions. The
resonator consists of an array of nanofluidic channels forming a
third order DFB Bragg grating resonator. The laser is realized
using a flexible fabrication technique, combining electron beam
lithography and UV lithography in a single polymer film, followed
by adhesive polymer wafer bonding.  The laser is straight-forward
to integrate on Lab-on-a-Chip micro-systems, e.g. for integrated
sensors, where coherent, tunable light in the visible range is
desired.

The authors thank S. Balslev and N. A. Mortensen for fruitful
discussions on the work presented in this Letter.

\begin{table}[htb]
\centering \caption{\label{tab:summary}Summary of the results. For
R6G solutions of larger refractive indices $n$, the laser
wavelength $\lambda$ increases, and the threshold pump fluence
$Q_{\rm{th}}$ and coupling loss $\gamma$ decrease.}
\begin{tabular}{c c c c c}
  Solvent & $n$ & $\lambda$/nm  & $Q_{\rm{th}}/\frac{\upmu\rm{J}}{{\rm{mm}}^2}$ & $\gamma$ \\
\hline
 Ethanol & 1.33 & 581.41 & 65 & .191 \\
 Ethylene glycol & 1.43 & 582.72 & 40 & .089 \\
 Ethylene glycol/ & 1.485 & 587.86 & 10 & .043 \\
 benzyl alcohol & & & & \\ \hline
\end{tabular}
\end{table}

\end{document}